\documentclass[prb,twocolumn,amssymb,amsmath,nobibnotes]{revtex4}
\usepackage{graphicx,color}% Include figure files
\usepackage{subfigure}
\usepackage{dcolumn}% Align table columns on decimal point
\usepackage{bm}% bold math
\usepackage{threeparttable}
\begin{document}

\newenvironment{figurehere}
  {\def\@captype{figure}}
  {}

\title{Switching and Rectification of a Single Light-sensitive Diarylethene Molecule Sandwiched between Graphene Nanoribbons}
\author{Yongqing Cai$^1$}
\author{Aihua Zhang$^1$}
\author{Chun Zhang$^{1,2}$}
\email{phyzc@nus.edu.sg}
\author{Yuan Ping Feng$^1$}

\affiliation{
	$^1$Department of Physics, National University of Singapore, 2 Science Drive 3, 117542 (Singapore)\\
    $^2$Department of Chemistry, National University of Singapore, 3 Science Drive 3,
117543 (Singapore)
}
\date{\today}

\begin{abstract}
The `open' and `closed' isomers of the diarylethene molecule that can be converted between each other upon photo-excitation are found to have drastically different current-voltage characteristics when sandwiched between two graphene nanoribbons (GNRs). More importantly, when one GNR is metallic and another one is semiconducting, strong rectification behavior of the `closed' diarylethene isomer with the rectification ratio $>$10$^{3}$ is observed. The surprisingly high rectification ratio originates from the band gap of GNR and the bias-dependent variation of the lowest unoccupied molecular orbital (LUMO) of the diarylethene molecule, the combination of which completely shuts off the current at positive biases. Results presented in this paper may form the basis for a new class of molecular electronic devices.

\end{abstract}

\maketitle

\section{INTRODUCTION}
As the trend of miniaturization continues, the next generation of electronic devices will almost undoubtedly be built on top of single molecules. Among all kinds of single-molecule based electronic devices, molecular switch and rectifier are probably most intensively investigated. In recent decades, lots of high-performance molecular switches with ON/OFF ratio $>$1000 have been proposed both experimentally and theoretically.\cite{Dulic_PRL,ChenY_APL,ZhuangM_PRB} On the other hand, the situation for single-molecule based rectifiers is not promising. Previously suggested molecular rectifiers usually have a rectification ratio in an order of 10,\cite{MetzgerRM_CP,StokbroK_JACS} and recently, it has been concluded that the rectification ratio originating from currently accepted rectification mechanisms of single-molecule based devices generally has a limit of 100.\cite{ArmstrongN_NL} The search of new mechanisms of molecular junctions that can exceed the limit of rectification ratio, 100, is one of central issues of the field.\cite{AndrewsDQ_JACS}
 
Photochromic molecules, which are capable of reversible photo transformations between two or more thermally stable states, have attracted remarkable interests in last decades due to their potential applications in light controllable molecular switches and memory devices.\cite{ZhangC_PRL,KawataS_CR,BrowneWR_ARPC,Del ValleM_NN} Diarylethene molecule is one of the most promising photochromic molecules for real applications due to its excellent thermal stability and high fatigue resistance. Metal-diarylethene junctions have been extensively studied both theoretically and experimentally for their applications in different molecular devices including switches and rectifiers.\cite{KudernacT_CC,IkedaM_CC,van der MolenSJ_NL} The key idea underlying the diarylethene based switches is that two light convertible isomers of the molecule, the `open' and the `closed' isomers (as shown in Fig. 1), have dramatically different conductances when connected with metal leads. As to the diarylethene-based rectifiers, the rectification behavior of the `open' isomer (rectification ratio in the order of 10) has been observed, and the origin of such rectification behavior was reported to be the large assymetric junction barrier at contacts formed between diarylethene-based polymer and metal leads.\cite{KawaiT_CAP} In last decades, great research effort has been paid to the tuning of the thermal stability, current ratio between `ON' and `OFF' states, and photochemical reversibility of the diarylethene molecule when mounted to the metal surface via modifying the core, spacer, and the linker group of the molecule(Fig. 1).\cite{Dulic_PRL,NakamuraS_PPA,ZhuangM_JCP,IrieM_CR,LiJ_PRL}

In this paper, we propose using graphene nanoribbons (GNRs), one of the most promising candidates for the next generation of electronic materials, to fabricate light controllable single diarylethene-molecular based switches and more importantly high-performance rectifiers. GNRs can be obtained either by tailoring two-dimensional (2-d) graphene or unzipping CNTs to one-dimensional (1-d) pattern.\cite{KosynkinDV_nature} GNRs have attracted tremendous amounts of interests due to their potential applications in nanoelectronics.\cite{ZhangLM_PRL,OssipovA_PRB} A GNR can be either semiconducting or metallic depending on its edge geometry. Ribbons with zigzag edges (zGNRs) were shown to be metallic if neglecting the magnetism, whereas the armchair edged ribbons (aGNRs) are semiconducting with energy gaps scaling with the inverse of the ribbon width.\cite{Castro NetoAG_RMP} Although it is well known that in 1-d and 2-d systems, the long-range magnetic order does not exist at finite temperature,\cite{MerminND_PRL,YanQ_CPL} magnetic property of graphene-based systems is a complex issue. In this paper, to simplify the problem, we stick to the non-magnetic case of GNRs.

GNR-diarylethene-GNR (G-D-G) junctions were considered in this paper. Via first principles method combining density functional theory (DFT) and nonequilibrium Green's functions' techniques (NEGF), we show that high performance molecular switches (ON/OFF ratio around 3000) and rectifiers with surprisingly high rectification ratio ($>$1000) could be obtained based on the G-D-G junctions. As above mentioned, previously proposed molecular rectifiers have a rectification ratio in an order of 10, and the search for high-performance single-molecule based rectifier is one of the central issues of the molecular electronics. Although for some single-molecule rectifiers, theoretical studies based on semi-empirical methods give high rectification ratio ($>>$100), more reliable first principles calculations for same systems predicted much lower ratio ($<<$100).\cite{AndrewsDQ_JACS} The rectification mechanism of G-D-G based rectifier described in this paper is different from those of previously proposed ones,\cite{AviramA_CPL,ElbingM_PNAS,StadlerR_AFM,LenfantS_NL,KornilovitchPE_PRB,KrzeminskiC_PRB,VuillaumeD_ME} and could form the basis for a new class of high performance single-molecule based light controllable devices.

\section{METHODS}
For the structural optimization and electronic properties, DFT calculations were performed with a plane wave basis set (cut-off energy 400 eV) and Perdew-Burke-Ernzerhof (PBE)-GGA approximation\cite{PerdewJ_PRL} via the use of VASP.\cite{KresseG_PRB} The structures were relaxed until the force is less than 0.02 eV/{\AA}. To ensure a junction with the central molecule free of strain, we also adjusted the molecule-lead separations by fixing the outermost layers of electrodes (Fig. 2a) and allowing other parts to relax to obtain the lowest-energy structure. The transport properties of the G-D-G junction were then calculated by the Atomistix ToolKit (ATK) code within the nonequilibrium Green's function (NEGF) formalism.\cite{BrandbygeM_PRB} The combination of DFT and NEGF techniques has proven to be powerful in qualitatively understanding transport properties of molecular junctions. Especially for cases of low bias voltages, the method is able to give results that quantitatively agree well with experiments.\cite{MarchenkovA_PRL,ZhangC_PRL} It is worthy mentioning here that since DFT often underestimates the energy gap of molecules, it may be problematic in transport calculations for systems far away from equilibrium state.    	 
In transport calculations, the aGNR (zGNR) electrode includes 2 (3) unit cells of the pristine GNR as shown in the shadowed area of Fig. 2a. 10 and 14 surface layers were included in contact regions of zGNR and aGNR junctions, respectively, which have been tested to be enough to screen the effects of the scattering region. Double-$\zeta$ polarized basis and a cutoff energy of 150 Ry for the grid integration within the Perdew-Burke-Ernzerhof (PBE)-GGA approximation were adopted in all the transport calculations.

\section{RESULTS}

\begin{figure}
\includegraphics[width=8.0cm]{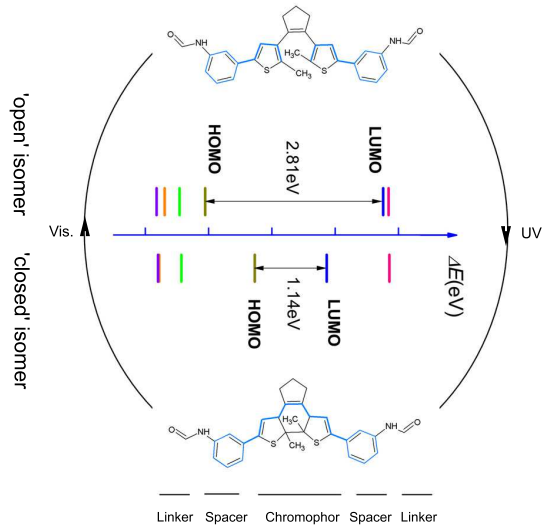}\\
\caption{(Color online) Molecular structures and photochemical interconversion of the `closed' and `open' diarylethene. The `closed' isomer (bottom) owning longer ¦Ð-conjugation length (blue line) shows larger conductance than the `open' isomer (upper). The molecules can be generally divided into `linker', `spacer', and `chromophore' groups. The calculated HOMO/LUMO gap values are in accordance with experimental observation of a blue or red shift in photo absorption spectra of the compounds Ref.\onlinecite{IkedaM_CC}.}
\label{Fig.1}
\end{figure}

In current study, a NHCO linker group is used in the diarylethene molecule as shown in Fig. 1. Calculated molecular energy levels of two isomers of the molecule are also shown in the figure. In supporting information, we show the isosurface of the highest occupied molecular orbital (HOMO) and LUMO orbitals for both isomers (Fig. S1). The molecule was connected to two GNR electrodes via forming hexagon and pentagon groups (Fig. 2a).\cite{Del ValleM_NN} According to convention, we refer to an armchair GNR with Na dimer lines as Na-aGNR and a zigzag GNR with Nz zigzag chains as Nz-zGNR (Fig. 2a) where Na and Nz denote the width of armchair and zigzag ribbons respectively. The band gap of an aGNR as a function of the ribbon width Na, $\vartriangle$(Na), exhibits three distinct families with $\vartriangle$(3$p$+1) $>$ $\vartriangle$(3$p$) $>$ $\vartriangle$(3$p$+2) (where $p$ is a positive integer), and for each family, the band gap is inversely proportional to the ribbon width. \cite{SonY_PRL,LongMQ_JACS} We chose 10-aGNR, 12-aGNR and 14-aGNR as representatives for all these 3 families. The 6-zGNR was used as an example of zGNRs.

Band structures of 4 different GNRs from our calculations are shown in Fig. 2b. The calculated energy gaps for armchair ribbons are 1.21, 0.54, 0.22 eV for 10-aGNR, 12-aGNR and 14-aGNR respectively, which are consistent with previous calculations.\cite{SonY_PRL} Both symmetrical x-GNR-diarylethene-x-GNR(x=z, a) and asymmetrical 6-zGNR-diarylethene-Na-aGNR (Na =10, 12, 14) junctions are investigated in this paper.

The conductance spectra of different symmetric GNR-diarylethene-GNR junctions at zero bias are shown in Fig. 3a and 3b. The `closed' isomer of diarylethene presents an overall much larger conductance around Fermi energy (EF) than the `open' isomer regardless of the electrode types, suggesting potential applications of this type of junctions in future design of light-driven molecular switches. For 6-zGNR junctions (Fig. 3a), the conductance at EF of the `closed' isomer is 3 orders larger than that of the `open' isomer. In the cases of 14-aGNR junctions (Fig. 3b), transmission gaps for both isomers near the Fermi energy are consistent with calculated band gaps of 14-aGNR. In Fig. 3c, we show isosurface of transmission eigen channels at EF for the 6-zGNR junction. Compared with the `open' isomer, the better coupling between the molecule and two leads for the case of the `closed' isomer that leads to higher conductance can be clearly seen from the figure.

\begin{figure}
\includegraphics[width=8.0cm]{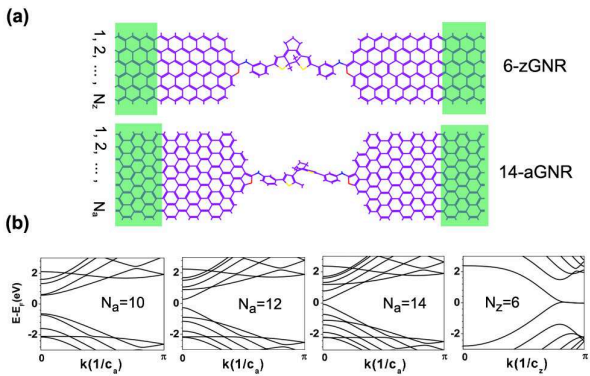}\\
\caption{(Color online) (a) Supercells of optimized 6-zGNR-`closed' diarylethene-6-zGNR and 14-aGNR-`open' diarylethene-14-aGNR junctions. Red, blue, violet and yellow lines denote bonds formed by O, N, C and S, respectively. Note that hydrogen atoms are not shown in the figure. The shadowed area denotes portions in the supercell chosen as electrodes. (b) Band structures of 10-, 12-, and 14-aGNR and 6-zGNR. The calculated band gaps are 1.21, 0.54, 0.22 eV for 10-aGNR, 12-aGNR and 14-aGNR respectively. Ca and Cz are the lattice constant of the unit cell of the aGNR and zGNR, respectively.}
\label{Fig.2}
\end{figure}

Since in reality, the molecular switch always operates under a finite bias, we calculated the current-voltage (I-V) characteristics of 6-zGNR-diarylethene-6-zGNR junction for two different isomers to further confirm the switching behavior. Results are shown in Fig. 4a. Compared with the `closed' isomer, the current tunneling through the `open' isomer is essentially zero at bias voltages ranging from -1.5 V to 1.5 V. The average `ON/OFF' current ratio is above 103 within the bias range under study. Switching behavior is also observed for aGNR junctions as shown in Fig. S2 in supporting information.

\begin{figure}
\includegraphics[width=8.0cm]{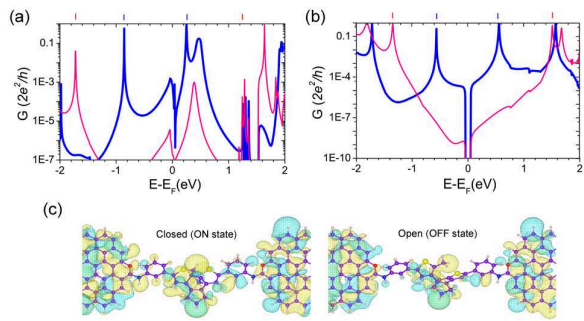}\\
\caption{(Color online) Conductance spectra at zero bias of 6-zGNR-diarylethene-6-zGNR (a) and 14-aGNR-diarylethene-14-aGNR (b) junctions with `closed' (blue) and `open' (red) isomers. The short lines above the plots show the HOMO/LUMO alignments with the electrode Fermi level. (c) Isosurface plot of transmission eigenstates at EF for `closed' (left) and `open' (right) isomers of diarylethene with 6-zGNR electrodes.}
\label{Fig.3}
\end{figure}

Interestingly, the current of the `closed' isomer varies with the bias voltage in a complicated manner. Starting from -0.4 V and +0.7 V, the current decreases quickly as the bias voltage increases, leading to significant negative differential resistances (NDRs) at these bias voltages. To understand the complicated behavior of the I-V curve of the `closed' diarylethene isomer, in Fig. 4b, we plot the variation of the logarithmic transmission as a function of single-electron energy and bias voltage, logT(e,V), for the 6-zGNR junction with the `closed' diarylethene isomer. Energy levels of molecular orbitals (HOMO-1, HOMO, and LUMO, LUMO+1) of the `closed' diarylethene isomer under different voltages, which are calculated by diagonalizing the molecule projected self-consistent Hamiltonian (MPSH),\cite{StokbroK_CMS} are also shown in the figure. We can see that within the bias range considered in this study, only the LUMO orbital of the molecule is inside the bias window lying between the chemical potentials of two leads. Therefore, only the LUMO orbital contributes to the total current. At zero bias, the LUMO orbital is about 0.2 eV above EF, and starts to enter the bias window around ¡À0.4 V at which the Fermi level of the source electrode align well with the LUMO orbital, leading to peaks in the I-V curve at these bias voltages. Contributions of interface states (ISs) localized at contacts between the molecule and zGNRs to the transmission can also be seen in the figure. Interface states can be clearly identified in the spatially resolved local density of states (LDOS) along the x axis (transport axis) as a function of energy and the fractional length plotted in Fig. 4c. Here, the fractional length is defined in one supercell (Fig. 2), and the LDOS along the x axis is calculated by summing up LDOS in y-z plane. At zero bias, ISs are formed at both source and drain sides lying 0.2 eV above LUMO level. As a positive bias voltage is applied, the IS at the drain side is shifted downward tracking the change of chemical potential of the drain. This interface state passes through the LUMO orbital at the bias voltage around 0.7 V, leading to a peak in I-V curve around this bias voltage. As the bias further increased, the interface state is more and more decoupled from LUMO, and the transmission through this state becomes lesser and lesser, resulting in the decrease of the current when the bias voltage is bigger than 0.7 V and NDR.

\begin{figure}
\includegraphics[width=8.0cm]{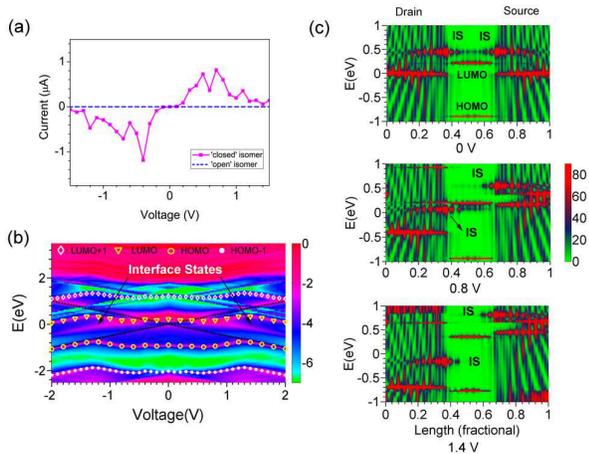}\\
\caption{(Color online) (a) I-V characteristics of 6-zGNR junctions with `closed' and `open' diarylethene. (b) Variation of logarithmic transmission as a function of energy and voltage, log T (e,V), for 6-zGNR symmetrical junction with `closed' diarylethene; Two straight black dotted lines denote Fermi energies of two electrodes. Contributions from HOMO-1 (white circles), HOMO (yellow triangles), LUMO (yellow circles) and LUMO+1 (white diamonds) are highlighted. (c) Spatially resolved local density of states (LDOS) along the transport axis as a function of energy and the fractional length at different bias voltages. The fractional length is defined in one supercell as shown in Fig. 2.}
\label{Fig.4}
\end{figure}

We then studied the coherent electron transport through an asymmetric junction consisting of a metallic 6-zGNR and a semiconducting 10-aGNR electrode (Fig. 5a). There have been several papers in literature studying electron transport through molecular junctions with semiconductor electrodes,\cite{ReuterMG_JPCA,MujicaV_CP,MujicaV_CP} where effects of the semiconducting gap on molecular rectification were discussed, but no significant rectification ratio was reported. In this paper, we focus on the G-D-G junction with the `closed' isomer of diarylenthene molecule since the `open' isomer gives essentially zero current in almost all cases compared with the `closed' one. In Fig. 5b, we show the variation of MPSH levels of frontier orbitals of the `closed' isomer as bias voltage changes from -1.5 V to 2 V. Two straight black lines in the figure denote Fermi energies of two electrodes. The difference between two Fermi energies defines the bias window that contributes to total current. The shadowed area shows the sweeping of the band gap of the 10-aGNR electrode under different voltages. As before, only the LUMO orbital enters the bias window within the bias range under study, therefore only this orbital has contributions to the overall current. In this case, the semiconducting a-GNR electrode presents another important constrain to electron tunneling: Any orbital inside the band gap does not contribute to transport. For positive biases, only when the bias voltage is bigger than 1 V, the LUMO orbital starts to contribute to electron transport, since only at these bias voltages, it is inside the bias window lying between two Fermi energies, and at the same time, outside the band gap. For negative biases, the LUMO orbital starts to have contributions to the current at -0.4 V for the same reason. Such behavior of the LUMO orbital strongly suggests asymmetric I-V characteristics that may lead to rectification effects. Indeed, the I-V curve showed in Fig. 5c clearly demonstrates the strong rectification behavior of this junction: The current is essentially zero for positive biases ranging from 0 to 1 V, and for negative biases, a significant current starts to occur at -0.4 V. The working bias range of the proposed rectifier is therefore from 0.4 to 1 V. In the inset of Fig. 5c, we show the rectification ratio of the rectifier as a function of bias within the working bias range, which is defined as the ratio between the current in forward direction and the current in reverse direction at a certain bias. The average rectification ratio is found to be around 1635 and a maximum value of $~$7500 is obtained around 0.5V. As aforementioned, previously experimentally or theoretically reported single-molecule based rectifiers generally have a typical rectification ratio $<<$100.\cite{MetzgerRM_CP,StokbroK_JACS,ArmstrongN_NL} These molecular rectifiers are normally based on metal-molecule-metal junctions, and the rectification behaviors are caused by the following mechanisms: the donor-acceptor mechanism,\cite{YanQ_CPL,AviramA_CPL,ElbingM_PNAS} asymmetric alignment of the frontier orbitals with the Fermi levels of the electrodes,\cite{StadlerR_AFM} and the different couplings between the molecule and two electrodes.\cite{LenfantS_NL,KornilovitchPE_PRB,KrzeminskiC_PRB}

\begin{figure}
\includegraphics[width=8.0cm]{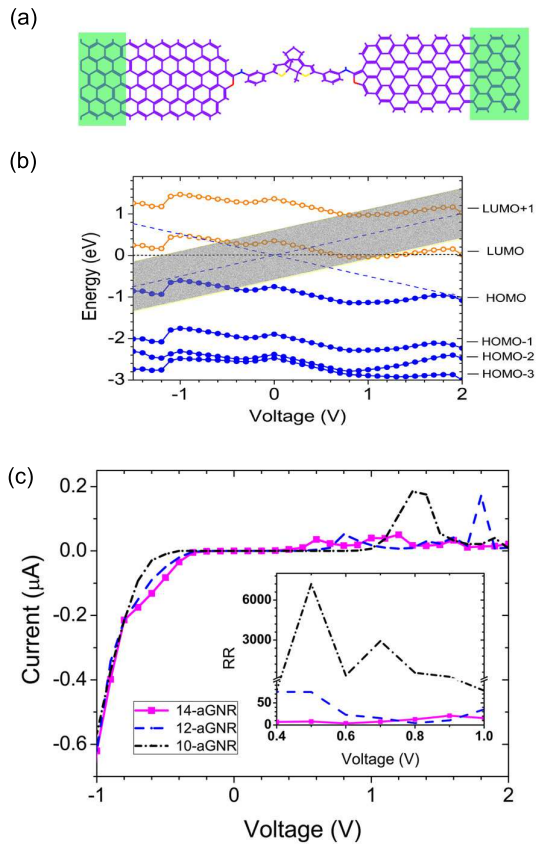}\\
\caption{(Color online)(a) Optimized atomic configuration of the 6-zGNR-diarylethene (`closed')-10-aGNR junction. (b) Evolution of MPSH levels as a function of voltage for the 6-zGNR-diarylethene-10-aGNR junction with the `closed' diarylethene. Blue dotted lines indicate Fermi energy levels of two electrodes. The shadowed area depicts the movement of the band gap of 10-aGNR under biases. Note that the LUMO is pinned to the Fermi level of the 6-zGNR (source) electrode from -0.3 to -1 V.  (c) I-V characteristics of 6-zGNR-(`closed') diarylethene-x-aGNR (x=10, 12 and 14) junctions. The inset shows the rectification ratio (RR) of these junctions from 0.4 to 1.0 V.}
\label{Fig.5}
\end{figure}

The origin of the high rectification ratio observed in the assymetric G-D-G junction is different from above-mentioned mechanisms. It comes from the the band gap of 10-aGNR and the bias-dependent variation of LUMO orbital of the diarylethene molecule, the combination of which completely shuts off the current at positive biases. To see more clearly the effects of the aGNR's band gap on the rectification behavior, we calculated the I-V characteristics of 6-zGNR-diarylethene-12-aGNR and 6-zGNR-diarylethene-14-aGNR junctions. As shown in Fig. 5c, both these junctions also show significant rectification behaviour, however, the rectification ratios are smaller (in the order of 10 and comparable to the metal-molecule-metal based rectifier \cite{ArmstrongN_NL,ElbingM_PNAS}) due to much smaller band gaps of 12- and 14 -aGNRs. These results suggest that the band gap is essential for the rectification behavior in G-D-G junctions, and the tuning of the band gap of GNRs may serve as an effective way to control the performance of assymetric GNR-diarylethene based rectifiers. Recent theoretical and experimental studies have shown that the band gap of GNRs can be effectively tuned by introducing antidot lattices or disorders.\cite{LongMQ_JACS,PedersenTG_PRL,BaiJ_NN,HanMY_PRL} In particular, significant band gaps have been observed in large graphene sheets decorated by periodic holes with the neck width around 10 nm, \cite{BaiJ_NN} and big transport gaps ($>$5 eV) were seen in wide disordered GNRs with the width $>$20 nm. \cite{HanMY_PRL} These studies pave the way for the realization of high-performance GNR-diarylethene based molecular rectifiers as proposed in this paper.
To demonstrate that both asymmetric electrodes and the diarylethene molecule are essential for the high rectification ratio as reported, we made a direct junction between the 6-zGNR and 10-aGNR, and calculated I-V curve. Results are shown in Fig. S3 in supporting information from which we can see that no significant rectification effects are present without the diarylethene molecule. Similar test were also done for other molecular wires such as carbon monoatom chain with various length and polyphenylene polymers sandwiched between the 6-zGNR and 10-aGNR electrodes, and none of these molecular wires shows significant rectification behavior.

\begin{figure}
\includegraphics[width=8.0cm]{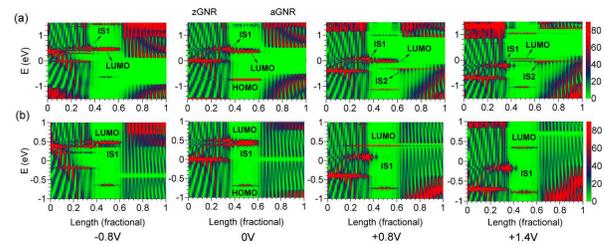}\\
\caption{(Color online)Spatially resolved local density of states along the transport axis for (a) 6-zGNR-`closed' diarylethene-10-aGNR (b) 6-zGNR-`closed' diarylethene-14-aGNR junctions at different bias voltages. The HOMO/LUMO of diarylethene localized in the middle region and interface states (ISs) localized at contacts can be clearly identified.}
\label{Fig.6}
\end{figure}

To further understand the detailed I-V characteristics of the 6-zGNR-diarylethene-aGNR junctions, we plot the spatially resolved LDOS of 10-aGNR and 14-aGNR junctions at different bias voltages in Fig. 6. The evolutions of HOMO and LUMO states with voltages are also shown in the figure. At zero bias, as we can see from the figure, interface states are only formed at the z-GNR side indicating different couplings between the molecule and two GNR electrodes. These interface states turned out to play important roles in electron transport through the junction. For 10-aGNR case, an interface state denoted as IS1 in the figure occurs on 6-zGNR side at zero bias. This state is strongly coupled to the LUMO state of the molecule due to their nearly degenerate energies, while, since both states are inside the band gap (close to conduction band) of 10-aGNR, they have no contributions to the current. As a positive bias is applied, both LUMO state and IS1 are shifted down towards the valence band of aGNR, and another interface state denoted as IS2 in the figure appears at aGNR side, as we can see from LDOS at +0.8 V. When the bias reaches +1.4 V, all three states, the LUMO, IS1, and IS2 are at the edge of valence band of aGNR. The strong coupling between the LUMO and two interface states at this bias significantly facilitates the electron transport, leading to the peak in I-V curve as shown in Fig. 5c. When a negative bias is applied, the IS1 and LUMO are shifted up and quickly enter the conduction band of aGNR as we can see from LDOS at -0.8 V. Within the bias range from -0.3 to -1 V, both LUMO and IS1 are pinned to the Fermi level of the source electrode as we can see from Fig. 5b and Fig. 6 due to the charge transfer from the source electrode to the molecule at these bias voltages as showed in Fig. S4 in supporting information. The nice alignment between LUMO and IS1 at these bias voltages leads to a significant increase of the current at these negative bias voltages compared to positive biases, and as a result, a high rectification ratio occurs.

For 14-aGNR case, the small current at fairly large bias ranging from 0 V to 2 V (as shown in Fig. 5c) can be understood by the decoupling of LUMO and the interface states. At zero bias, similar to the case of 10-aGNR, an interface state IS1 occurs at zGNR side, and this state has similar energy as the LUMO. When a positive bias is applied, the LUMO and IS1 are well separated in energy, and strongly localized in the center region and left contact, respectively, as we can see from LDOS at +0.8 V and +1.4 V. Therefore both of them have no significant contributions to electron transport, resulting in small current at a fairly large range of positive biases as shown in Fig. 5c. Detailed information of transmission as function of bias voltage and energy for 10-aGNR, 12-aGNR and 14-aGNR cases are shown in Fig. S5 and S6 in supporting information, from which we can clearly see the contribution of LUMO and interface states to electron transmission.

\section{CONCLUSIONS}
We have investigated the coherent electron transport through a single light sensitive diarylethene molecule sandwiched between two GNRs via first principles calculations based on DFT and NEGF techniques. The `open' and `closed' isomers of the diarylethene molecule that can be converted between each other upon photo-excitation were found to have drastically different current-voltage (I-V) characteristics, suggesting that the GNR-diarylethene-GNR junction can be used as a high-performance light controllable molecular switch with a current ratio of ON/OFF states above 103. More importantly, we reported the strong rectification behavior of 6-zGNR-diarylethene-x-aGNR junctions with average rectification ratios of 1635, 34 and 11 for cases of x=10, 12, and 14 respectively. The rectification mainly arises from the bias-dependent variation of the energy level of the diarylethene molecule's LUMO orbital relative to the band gap of the semiconducting GNR. Therefore, the appropriate choice of the band gap of the semiconducting electrode and the alignment of the molecule's frontier orbitals with the semiconducting gap are important for achieving the high rectification ratio. Moreover, the interface states localized in molecule-GNR contacts were found to play important roles in electron tunneling. The rectification mechanism discussed in this paper provides a new way to achieve high rectification ratio ($>>$100) in single-molecule based devices, and the asymmetric G-D-G junctions may form the basis for a new class of high-performance molecular rectifiers.

Acknowledgement: Computations were performed at the Centre for Computational Science and Engineering(CCSE) and High Performance Computing Centre at NUS.

\end{document}